\documentclass{moriond}

\def\be{\begin{equation}}
\def\ee{\end{equation}}
\def\bea{\begin{eqnarray}}
\def\eea{\end{eqnarray}}

\newcommand{\Photo}{\includegraphics[height=35mm]{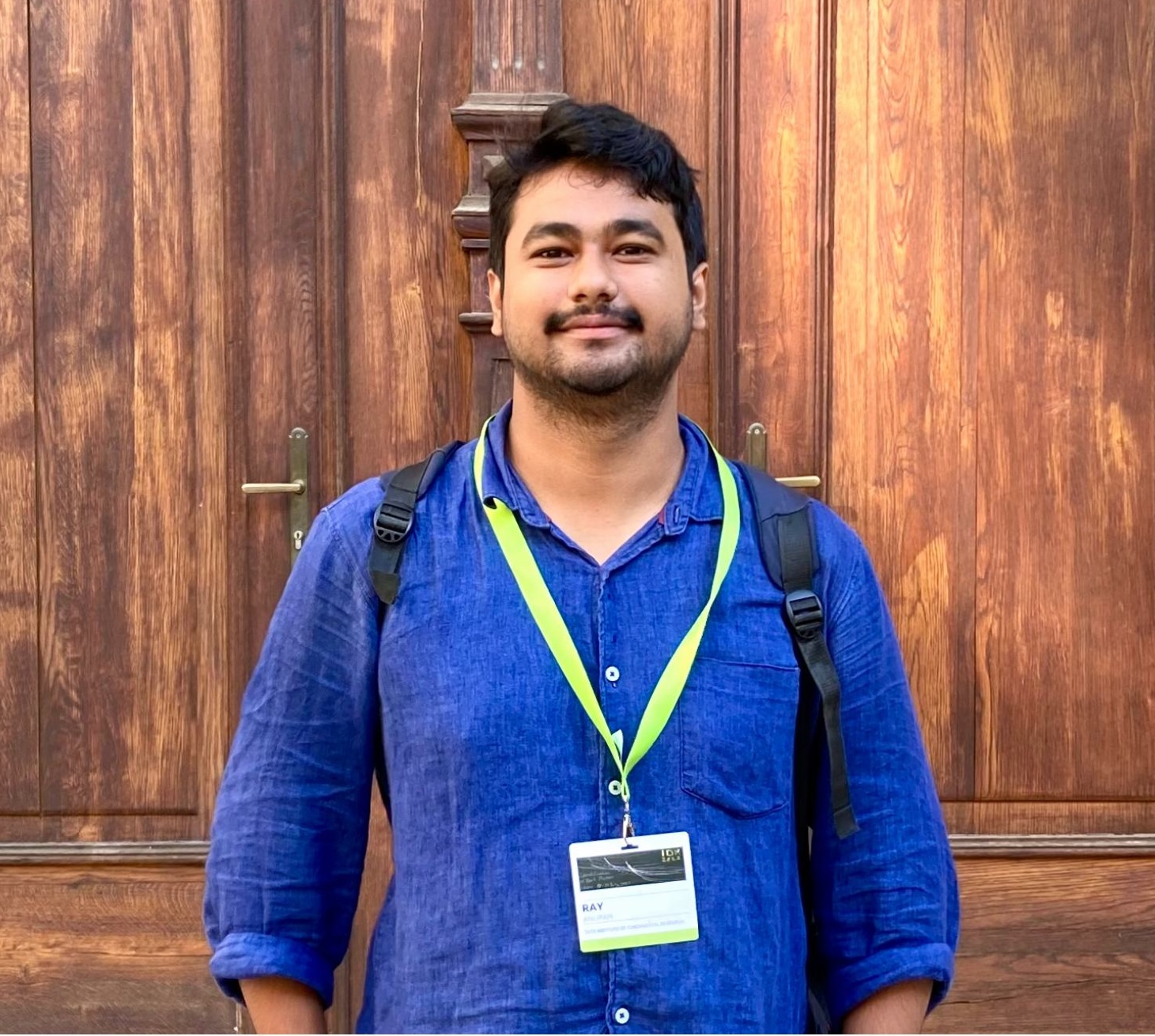}}

\begin{document}
\vspace*{4cm}
\title{Detecting Dark Matter with Neutron Stars}

\author{Anupam Ray}

\address{Department of Physics, University of California, \\ Berkeley, California 94720, USA}

\maketitle\abstracts{Neutron stars offer powerful astrophysical laboratories to probe the properties of dark matter. Gradual accumulation of heavy, non-annihilating dark matter in neutron stars can lead to the formation of comparable-mass black holes, and non-detection of gravitational waves from mergers of such low-mass black holes can constrain such dark matter interactions with nucleons. These constraints, though dependent on the currently uncertain binary neutron star merger rate density, are significantly more stringent than those from direct detection experiments and provide some of the strongest limits on heavy, non-annihilating dark matter interactions. Additionally, dark matter with baryon number–violating interactions can induce excess heating in cold neutron stars and is thus significantly constrained by thermal observations of cold neutron stars.}

\section{Introduction}
Cosmological observations offer compelling evidence for a non-baryonic form of matter, commonly referred to as dark matter (DM), as a major constituent of our Universe.  While its gravitational effects are well established $-$ from galaxy rotation curves to large-scale structure formation $-$ the non-gravitational interactions of DM remains one of the most profound open questions in modern physics. Despite extensive efforts across direct detection, indirect detection, and collider-based searches, no conclusive evidence for the non-gravitational interactions of DM has been observed~\cite{Cirelli:2024ssz}.\\
In this article, we demonstrate how neutron stars can serve as powerful astrophysical laboratories to probe the properties of dark matter. With their extreme densities and strong gravitational fields, neutron stars (NSs) offer a unique environment where even feeble interactions between DM and ordinary matter can leave observable imprints. We highlight how electromagnetic and gravitational wave observations of neutron stars can be used to explore a broad spectrum of dark matter models, spanning an extensive range of mass scales. We focus on two specific dark matter scenarios: heavy non-annihilating dark matter and WIMP-like dark matter with baryon-number-violating interactions, showing that observations of neutron stars can place leading constraints on both. The rest of the work is organized as follows: in Section 2, we highlight how GW observation of binary neutron stars can provide novel powerful constraints on heavy non-annihilating DM interactions, and in Section 3, we explore how thermal observation of cold neutron stars can set leading exclusions on baryon-number-violating processes.
\section{Searching Non-annihilating  dark matter with neutron stars}
Dark matter particles from the Galactic halo can interact with the constituents of stellar objects as they pass through them. Such interactions can cause the DM particles to lose kinetic energy and down-scatter to velocities below the local escape velocity, eventually becoming gravitationally bound to the stellar objects. Over time, this capture process enables celestial bodies to accumulate a significant population of DM within their interiors~\cite{1987ApJ...321..571G,Dasgupta:2019juq}. Neutron stars, due to their extraordinarily high baryonic densities, are particularly efficient at capturing DM particles, even when the dark matter–baryon interaction cross-section is exceedingly small, making them ideal astrophysical laboratories for probing DM interactions.\\
For heavy, non-annihilating dark matter, the captured particles gradually thermalize within the stellar interior through repeated collisions with stellar constituents. As they lose energy and reach thermal equilibrium, they eventually settle into a tiny core region, the size of which is governed by the balance between thermal pressure and the gravitational potential of the host star. Since a significant number of dark matter particles accumulate in this small region, the resulting dark matter core can reach extremely high densities. If the core mass exceeds a critical threshold, it becomes gravitationally unstable and undergoes collapse, potentially forming a micro black hole  at the stellar core. This nascent black hole, formed via dark core collapse, if not sufficiently light, can rapidly swallow the host, transmuting it into a black hole of comparable mass. Therefore, mere existence of celestial bodies, specifically old neutron stars, provides significant exclusion on heavy non-annihilating dark matter interactions~\cite{Kouvaris:2010jy,McDermott:2011jp,Dasgupta:2020dik,Bramante:2023djs}. In fact, these existence-inferred exclusions currently provide the world-leading constraints on such dark matter interactions, surpassing those from terrestrial direct detection experiments, especially in the high-mass regime where sensitivity is otherwise limited due to the smaller fluxes.\\
In the following, we argue that gravitational wave observations of neutron stars can also provide a novel and complementary probe of heavy non-annihilating DM interactions~\cite{Bhattacharya:2023stq}. The central idea is that continued accumulation of such DM particles in binary neutron stars can lead to the formation of anomalously low-mass black holes, with masses in the range of  $\sim (1 - 2.5)\,M_{\odot}$, and GWs from mergers of such low-mass black holes can be searched in the LIGO-Virgo-KAGRA (LVK) detector network. Given null detection so far, it sets strong exclusions on heavy non-annihilating DM interactions. These constraints  crucially depend on the binary neutron star merger rate density, which remains uncertain; however, if the true rate lies near the upper end of current estimates, the resulting constraints could offer unprecedented sensitivity to such DM interactions. Naturally, as the detectors continue to collect data, they will explore progressively larger regions of the relevant DM parameter space, uncovering a new promising science case for these gravitational wave detectors.\\
In order to determine the GW-inferred exclusions on heavy non-annihilating DM interactions, we first estimate the timescale for transmutation $(\tau_{\rm trans})$. This timescale consists of two primary phases:  a) the time required to form a micro black hole at the stellar core,  and b) the time required for it to destroy the host star, and it primarily depends on DM mass, DM-nucleon interaction cross-section, along with the ambient DM density. For a successful transmutation of the host, $\tau_{\rm trans}$ must be shorter than the available evolutionary time of the binary system, i.e., $\tau_{\rm trans} \leq t_0 -t_f$, where $t_f$ denotes the binary formation time and $t_0$ denotes the current age of our Universe. So, given a population of the host stars (binary neutron stars), the merger rate density of transmuted black holes~\cite{Bhattacharya:2023stq,Dasgupta:2020mqg}
\begin{equation}
	R_{\rm TBH} = \int dr \, \frac{df}{dr} \int_{t_*}^{t_0} dt_f \, \frac{dR_{\rm BNS}}{dt_f} \times \Theta\left[t_0 - t_f - \tau_{\rm trans}(m_\chi, \sigma_{\chi n}, \rho_{\rm ext}(r, t_0))\right]\,,
\end{equation}
where $R_{\rm BNS}$ denotes binary NS merger rate density and $\theta$ function accounts for successful transmutation fraction. The spatial distribution of the host stars, $df/dr$ is taken as a uniform 1d distribution in $r \in (0.01, 0.1)~\mathrm{kpc}$ where $r$ denotes the Galactocentric distance. The ambient DM density $(\rho_{\rm ext})$ is modeled using a Navarro–Frenk–White  profile; for a conservative estimate, we neglect its time evolution. The lower-limit of the $t_f$ integral  corresponds to redshift $z_* = 10$, marking the earliest time of binary star formation. $dR_{\rm BNS}/dt$ 
is the rate-density of binary NS mergers for a given formation time which is determined  by the cosmic star formation rate, the fraction of stellar mass in binaries, and the delay time distribution. The overall normalization of $R_{\rm BNS}$ is fairly uncertain and  is treated as a free parameter, varying within the range of (10 $-$1700) Gpc$^{-3}$\,yr$^{-1}$, as suggested by the recent LVK observations~\cite{KAGRA:2021duu}. For a more detailed estimate the transmuted merger rate, and  its dependence on various model parameters, see the original works~\cite{Dasgupta:2020mqg,Bhattacharya:2023stq}.\\
\begin{figure}
	\label{fig2}
	\includegraphics[width=0.95\textwidth]{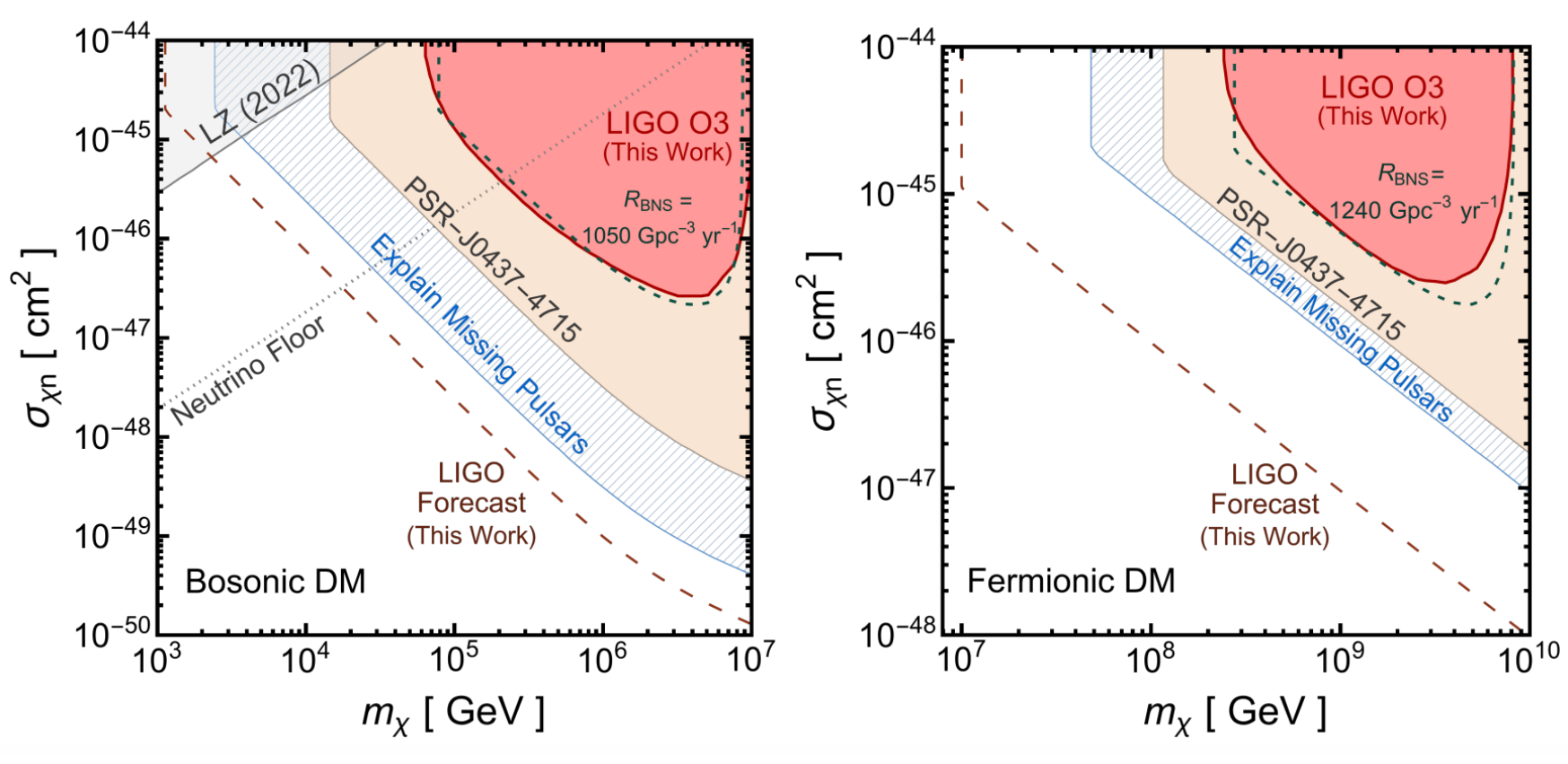}
	\caption{Gravitational-wave constraints on heavy non-annihilating dark matter interactions. The left panel shows results for bosonic dark matter, and the right panel for fermionic dark matter. In both panels, the pink shaded regions represent the 90\% credible Bayesian exclusion limits derived from the non-detection of gravitational waves originating from mergers of low-mass transmuted black holes in LVK data. The green dashed lines correspond to 90\%  upper limits obtained using a frequentist approach, assuming fixed values of $R_{\rm BNS} = 1050\,(1240)$\,Gpc$^{-3}$\,yr$^{-1}$, roughly matches the corresponding Bayesian limits. It is evident that constraints derived from neutron stars are significantly stronger than those from direct detection and can probe parameters well below the neutrino floor  (left panel). See text for a more detailed discussion on how these GW-inferred limits compare to the existing limits.}
\end{figure}
Given the non-detection of such low-mass transmuted black hole mergers in the LVK data, we derive novel exclusions on heavy non-annihilating DM interactions, as shown in Fig.~1 (which is taken from~\cite{Bhattacharya:2023stq}). For the statistical inference, we adopt a Bayesian likelihood framework, with complete methodological details available in the original work~\cite{Bhattacharya:2023stq}. In Fig.~1, the pink shaded regions labeled “LIGO O3” show the 90\% credible Bayesian exclusion limits obtained by marginalizing over the uncertain normalization parameter $R_{\rm BNS} \in (10 - 1700)$\,Gpc$^{-3}$\,yr$^{-1}$. Quantitatively, for bosonic DM (left), we find an exclusion limit of $\sigma_{\chi n} < 2.5 \times 10^{-47}~\mathrm{cm}^2$ for $m_{\chi} = 5$ PeV, and it weakens as $1/m^{3/2}_{\chi}$ at smaller masses up to 0.06 PeV. For fermionic DM (right), the exclusion limit is $\sigma_{\chi n} < 2.4 \times 10^{-46}~\mathrm{cm}^2$ at $m_\chi = 3.6 \times 10^3~\mathrm{PeV}$, weakening as $\sim 1/m_{\chi}$ up to 240
PeV. The exclusion limits for fermionic DM applies to much heavier DM masses because the collapse of a degenerate fermionic core is more strongly resisted by Fermi pressure, requiring a larger DM mass for successful transmutation. In Fig.~1, the dark green curves labeled ``$R_{\rm BNS} = 1050\,(1240)$\,Gpc$^{-3}$\,yr$^{-1}$" represent the 90\% frequentist exclusion limits obtained under the assumption of fixed binary neutron star merger rates. Our 90\% credible Bayesian limits are numerically comparable similar to these, enabling a meaningful comparison between the two approaches. We also note that for $R_{\rm BNS} = 10$\,Gpc$^{-3}$\,yr$^{-1}$ (the lowest value in the allowed range), no frequentist exclusion on the DM parameter space is obtained. The minimum values of $R_{\rm BNS}$ for which current data can start can constrain some of the DM parameter space in a frequentist analysis are approximately 900 (1110) \,Gpc$^{-3}$\,yr$^{-1}$  for bosonic (fermionic) DM.\\
In Fig.~1, we also present the leading constraint from underground direct detection experiments (gray shaded region labeled “LZ (2022)”)~\cite{LZ:2022lsv} for comparison. It is evident that the GW-inferred exclusions are significantly stronger than those from direct detection in the high-mass regime. This simply arises from the fact that neutron stars, owing to their much larger exposure compared to human-made underground detectors, are inherently far more sensitive to the extremely low fluxes of heavy dark matter. Furthermore, it is crucial to emphasize that these GW-inferred exclusions can probe parameter space well below the neutrino floor (as shown in the left panel of Fig.~1), where direct detection is hindered by the irreducible neutrino background, thus providing a unique advantage for GW detectors in this region.\\
We also compare the GW-inferred exclusions with the constraint derived from the existence of the nearby old pulsar PSR-J0437-4715 (beige shaded region in Fig.~1)~\cite{Kouvaris:2010jy,McDermott:2011jp}. Currently, our constraint, inferred from the existing LVK data, is weaker than the PSR-J0437-4715 constraint. However, due to the fundamentally different systematics between GW detection and radio pulsar searches, our result is complementary and has the potential to become the leading constraint with upcoming GW observations. For instance, the dashed curves labeled  “LIGO Forecast” represent the projected upper limits that could be achieved in the future if the exposure for the LVK detectors, $\langle V T \rangle$, increases by a factor of 50 $-$ a scenario that is plausible by the end of this decade. Additionally, with next-generation detectors such as the Einstein Telescope and Cosmic Explorer, the sensitivity can further improve by several orders of magnitude~\cite{Bhattacharya:2023stq}.\\
We note that both the GW-inferred and EM-inferred exclusions terminate at $m_{\chi} \geq 10^7~(10^{10})$\,GeV for bosonic (fermionic) dark matter. This is simply because, for such high DM masses, the nascent black hole formed via DM-induced collapse becomes increasingly lighter, causing Hawking evaporation to dominate over accretion. Consequently, the black hole evaporates before it can grow, resulting in an unsuccessful transmutation. However, potential modifications to Hawking evaporation due to quantum gravity effects, commonly known as Memory-burden effect, could significantly suppress evaporation rates, which in turn opens up the possibility of probing even higher dark matter masses, as showed in~\cite{Basumatary:2024uwo}. For discussions on the transmutation of non-compact celestial bodies and related phenomenological signatures, we refer the reader to Refs.~\cite{Ray:2023auh,Bhattacharya:2024pmp}.
\section{Probing Baryon-number-violating interactions with neutron stars}
\begin{figure}
	\label{fig3}
	\includegraphics[width=0.445\textwidth]{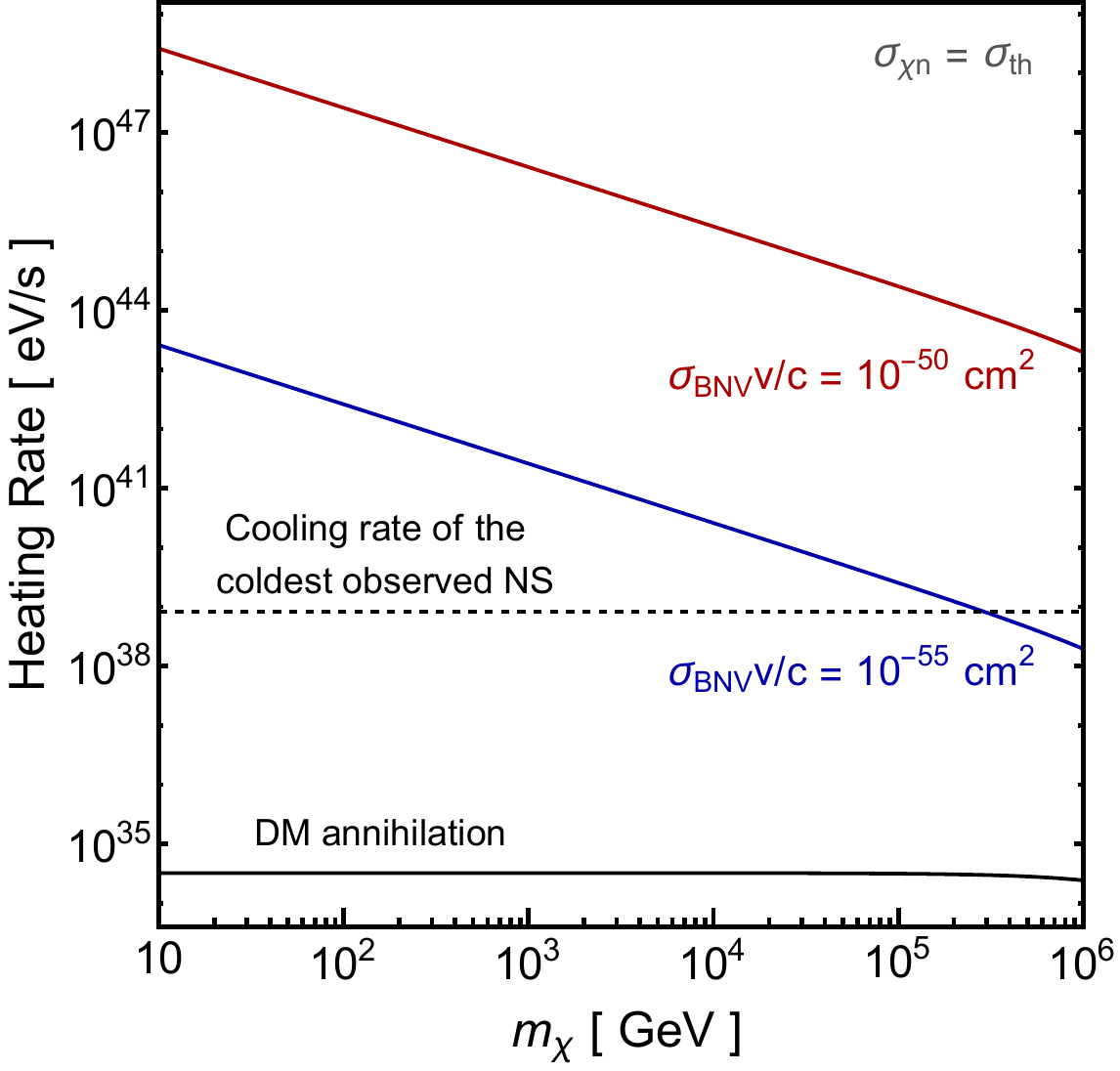}
	\hspace*{0.75 cm}
	\includegraphics[width=0.47\textwidth]{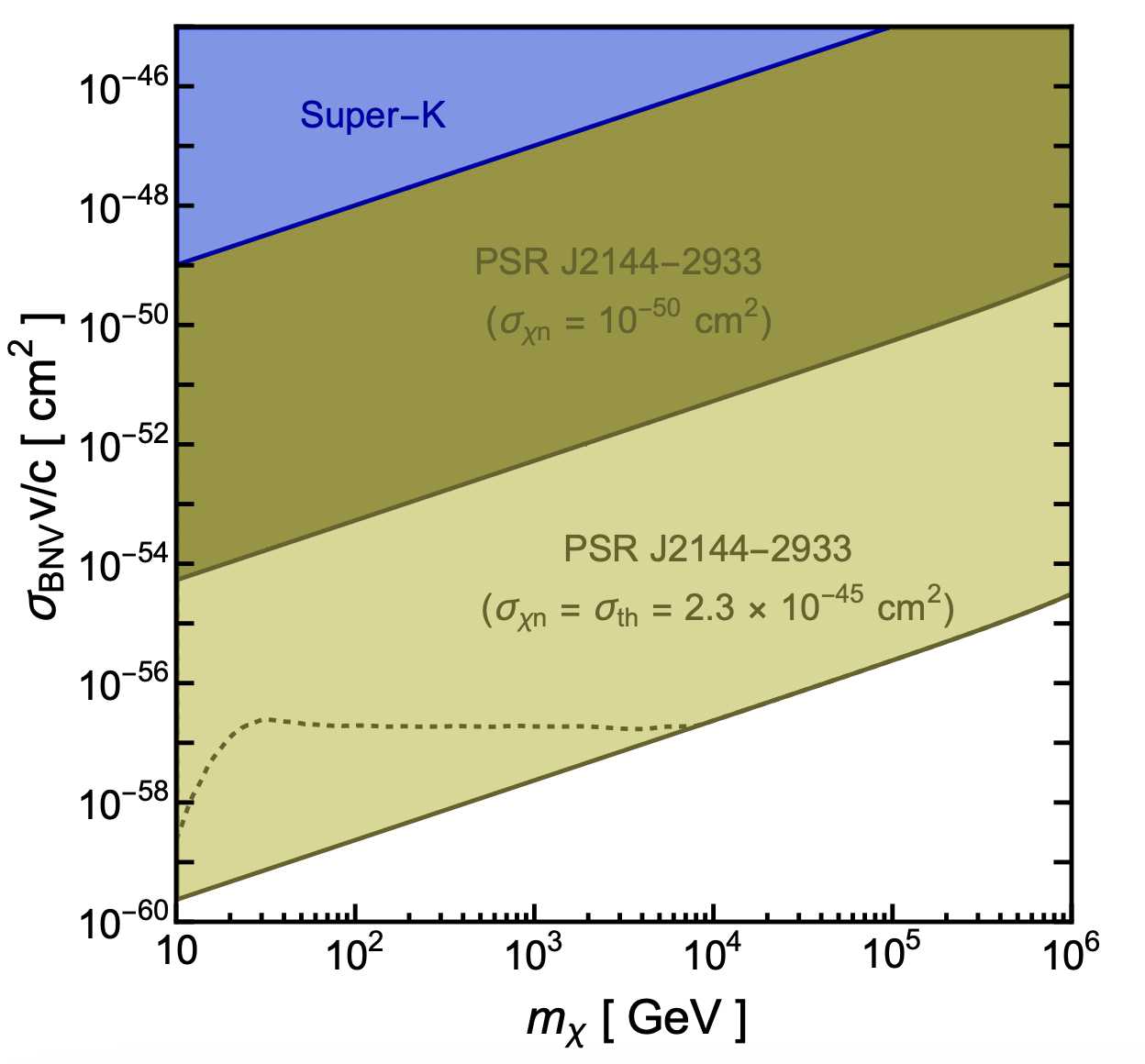}
	\caption{(Left) Heating rates from the DM-catalyzed baryon destruction process with $v \sigma_{\rm{BNV}}/c = 10^{-50}$ cm$^2$ (red) and $10^{-55}$ cm$^2$ (blue). DM-nucleon scattering cross-section is set at $\sigma_{\rm{th}}$ to ensure maximal capture of $\chi_1$ particles within the NS. The cooling rate of the coldest observed NS, PSR J2144-3933, is also shown for comparison. It is evident that heating rates induced by $\Delta B \neq 0$ processes are significantly higher than those from DM annihilation scenarios (dashed black), which are currently not excluded. (Right) Exclusions on BNV interactions s from the non-observation of anomalous heating in PSR J2144-3933 (yellow shaded regions). See text for more details.}
\end{figure}
Neutron stars, owing to their extremely high baryon densities, serve as exceptional laboratories for probing baryon-number-violating (BNV) interactions involving DM particles~\cite{Ema:2024wqr}. In the following, we demonstrate how electromagnetic observations of cold neutron stars can be leveraged to place world-leading constraints on such BNV interactions. For concreteness, we consider two dark matter species, $\chi_1$ and $\chi_2$, with nearly degenerate masses and $m_{\chi_2} > m_{\chi_1}$. Due to baryon-number-violating interactions, $\chi_1$ can destroy nucleons via $\chi_1 + p/n \to \chi_2 + e^+/\bar{\nu}$, with the resulting Standard Model particles rapidly absorbed within the neutron star, causing excessive heat generation. For this heating to be efficient, $\chi_2$ must be recycled back to $\chi_1$ on a relatively short timescale. This recycling can occur either through oscillations induced by small mass splittings and mixing between the two states, or via direct decay of $\chi_2$ back to $\chi_1$.\\
We begin by estimating the amount of heat generated through this mechanism. This requires first calculating the total number of dark matter particles $(\chi_1)$ captured by the neutron star over its lifetime $(t_{\star})$. If the mass of $\chi_1$, $m_{\chi_1}$, falls within the GeV–PeV range, they can be trapped within the neutron star after a single collision, and by assuming  they do not annihilate among themselves, the number of such particles in the neutron star is~\cite{Bhattacharya:2023stq}
\begin{equation}
	N_{\chi_1} = \epsilon_{\rm{cap}} \, \sqrt{\frac{6}{\pi}} \, \frac{\rho_{\chi_1}}{m_{\chi_1}} \, \pi R_\star^2 \, \bar{v} \frac{v_{\rm esc}^2}{\bar{v}^2}\left(1 - \frac{1 - e^{-A^2}}{A^2} \right) t_\star\,,
\end{equation}
where $\epsilon_{\rm{cap}} = \min\left(1, \frac{\sigma_{\chi n}}{\sigma_{\rm{th}}}\right)$ is the capture efficiency, which depends on the DM-nucleon scattering cross section $\sigma_{\chi n}$, with $\sigma_{\rm{th}} = \pi R_\star^2 / N_n$ denoting the critical threshold cross section above which all of the transiting $\chi_1$ get trapped. $\bar{v}$ is the mean velocity of $\chi_1$ in the Galactic halo, $\rho_{\chi_1}$ is the ambient DM density of $\chi_1$, $v_{\rm esc}$ is the surface escape velocity of the NS, and $R_{\star}$ is the radius of the NS. The dimensionless factor involving $A^2 = \frac{6 m_\chi m_n v_{\rm{esc}}^2}{\bar{v}^2 (m_\chi - m_n)^2}$, with $m_n$ being the nucleon mass, accounts for inefficient momentum transfer in the DM-nucleon scattering and evaluates to unity in the parameter range of interest $(m_{\chi_1} \leq 10^6\,\rm{GeV})$.\\
Therefore, the heating rate due to neutron destruction is~\cite{Ema:2024wqr}
\begin{equation}
\frac{dE_{\rm{heat}}}{dt} = N_{\chi_1} \times v \sigma_{\rm{BNV}} \times n_n m_n
\end{equation}
where $n_n$ denotes the neutron number density inside the NS, assumed to be uniform, and $v\sigma_{\rm{BNV}}$ denotes the baryon-number-violating cross section times the relative velocity for $\Delta B \neq 0$ process, treated as a free parameter.  For the coldest NS observed till date, PSR J2144-3933, which has a surface temperature of 2.85 eV, the heating rate is found to be $\sim 2.6 \times 10^{47}$ eV/s for $v\sigma_{\rm{BNV}}/c = 10^{-50} \, \rm{cm}^2$. This exceeds the cooling rate of this particular NS,  which is $6.4 \times 10^{38}$ eV/s, implying that such BNV cross-sections would result in excessive heating and are thus strongly constrained.\\
In Fig.~2 (left panel) (which is taken from~\cite{Ema:2024wqr}), we show the heating rate resulting from the above process for two benchmark values of the BNV cross sections. The DM-nucleon interaction cross section is taken to be at the critical threshold value to ensure maximal capture of the $\chi_1$ particles. As evident, the resulting heating rate is significantly higher than the cooling rate of PSR J2144-3933, thereby excluding such benchmark BNV cross sections. Here it is important to emphasize that heating rate arising from captured DM annihilation (dashed black) is significantly lower as compared to the BNV processes, and such annihilation scenarios currently remain unconstrained by observations.\\
In Fig.~2 (right panel), we also show the constraints on BNV interactions from the non-observation of any anomalous heating in PSR J2144-3933 (yellow shaded regions) for some benchmark DM-nucleon scattering cross sections. The dashed yellow curve corresponds to DM-nucleon scattering cross sections that are currently allowed by direct detection searches, indicating that even very small values of BNV cross sections are excluded. It is evident that electromagnetic observations of cold neutron stars can provide leading constraints on $\Delta B \neq 0$ processes, significantly stronger than those obtained from proton decay searches at Super-Kamiokande (blue shaded region).
\section*{Acknowledgments}
I wish to thank all of my collaborators  for valuable contributions in the original works. I also acknowledge support from the National Science Foundation (Grant No. PHY-2020275), and the Heising-Simons Foundation (Grant 2017-228).
\section*{References}
\bibliography{moriond}

\end{document}